# The effect of material defects on resonant spin wave modes in a nanomagnet

Md Ahsanul Abeed[1], Sourav Sahoo[2], David Winters[1], Anjan Barman[2*] and Supriyo Bandyopadhyay[1†]

[1]Dept. of Electrical and Computer Engineering, Virginia Commonwealth University, Richmond, VA 23284, USA

[2]Dept. of Condensed Matter Physics and Materials Science, S. N. Bose National Center for Basic Sciences, Kolkata 700 116, INDIA

**Abstract**

We have theoretically studied how resonant spin wave modes in an elliptical nanomagnet are affected by fabrication defects, such as small local thickness variations. Our results indicate that defects of this nature, which can easily result from the fabrication process, or are sometimes deliberately introduced during the fabrication process, will significantly alter the frequencies, magnetic field dependence of the frequencies, and the power and phase profiles of the resonant spin wave modes. They can also spawn new resonant modes and quench existing ones. All this has important ramifications for multi-device circuits based on spin waves, such as phase locked oscillators for neuromorphic computing, where the device-to-device variability caused by defects can be inhibitory.

[*]Correspondence to abarman@bose.res.in
[†] Correspondence to sbandy@vcu.edu

# I. Introduction

Spin wave modes in a nanomagnet govern its dynamical behavior. Magnetization reversal in a nanomagnet – a phenomenon that undergirds the operation of virtually all magnetic switching devices used in memory, logic, and non-Boolean computing – is mediated by two major types of collective excitations: spin waves and domain wall motion[1]. It is therefore imperative to understand how spin wave excitations behave in a nanomagnet since it can determine the speed of magnetic reversal and some other properties such as the switching error rate. Today, spin waves have an even more important role to play since they are the central ingredients in spin wave logic[2,3], magnetic nano-oscillators for nanoscale microwave generators[4] and certain types of artificial neural networks[5].

Spin wave modes in nanomagnets have been a subject of research for about two decades[6]. Initially, interest was focused on understanding their behavior and damping inside individual nanomagnets and their dependence on size, shape and initial magnetic configuration[7-11]. Later, interest shifted towards understanding the collective dynamics in an array of nanomagnets[12,13], which emerged as a potential candidate for a two-dimensional magnonic crystal. Much work has been reported on the dependence of spin wave dynamics on the size and the shape of nanomagnets, the lattice constant, the lattice symmetry of the array, bias field strength and orientation, as well as binary and bi-component nature[14-20]. Important physical phenomena such as spin wave mode splitting, mode cross-over, dynamic dephasing, transition between various collective regimes, intrinsic and extrinsic dynamical configurational anisotropy, etc. have been reported. Energy-efficient spin-wave device concepts have also been proposed based on shaped nanomagnet arrays[21]. Many of these studies have considered edge roughness and deformation of the nanomagnets and their effect on spin waves, but the influence of various types of defects in nanomagnets on spin wave modes and dynamics has remained unexplored.

In this paper, we have studied theoretically how the resonant spin wave modes in an elliptical nanomagnet are affected by different types of imperfections or "defects". Specifically, we have studied how the power and phase profiles of these modes are altered by small local variations in the nanomagnet thickness. Our study reveals that defects (various types of local thickness variations) can have significant effects on the spin wave modes. They can change the frequencies and phases of the resonant modes. That can have serious consequences for spin-wave based circuits comprising many devices where the device-to-device variability caused by defects can impair circuit functionality. An example of this is phase locked oscillators used for neuromorphic computation[22] where the device-to-device variability may pose a challenge. Defects can also generate completely new modes or quench existing ones. That can have consequences in many other applications as well, e.g. spin wave logic.

Defects (thickness variations) are usually unavoidable during the fabrication process. Fig. 1 shows atomic force micrographs of nanomagnets fabricated in our lab with standard electron beam evaporation of Co into lithographically delineated windows opened in e-beam resist. The resists are patterned with electron beam lithography and the nanomagnets are produced by lift-off. The lateral dimensions of these nanomagnets are on the order of 100-300 nm and their thickness is on the order of 16 nm. Note that in Fig. 1, the nanomagnets have "rims", i.e. the thickness is much larger along the periphery than at the center. This type of defect is an aftermath of the lift-off process and is fairly common.

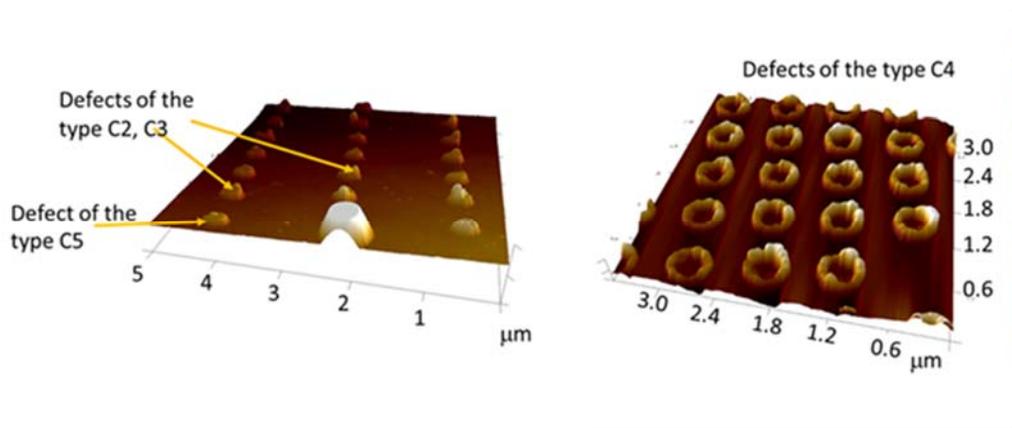

Fig. 1: Atomic force micrographs of arrays of Co nanomagnets deposited on a substrate using electron beam lithography, electron beam evaporation of Co on to the patterned substrate, followed by lift-off. The nanomagnets have various defects such as thickness variation along the plane (classified as defects of type C2 and C3), a raised region in the center (classified as defect C5) and cratering or larger thickness along the periphery (classified as defect C4). Reproduced from ref. [30] with permission of the American Physical Society © American Physical Society.

Other types of defects will involve "voids" (missing material) or "mesas" (excess material) at certain locations on the nanomagnet's plane. They are usually caused during metal evaporation. The types of defects that we have studied in this work are shown schematically in Fig. 2. Note that defect type C4 approximates the structure shown in the right panel of Fig. 1, while the other types are representative of the defects shown in the left panel.

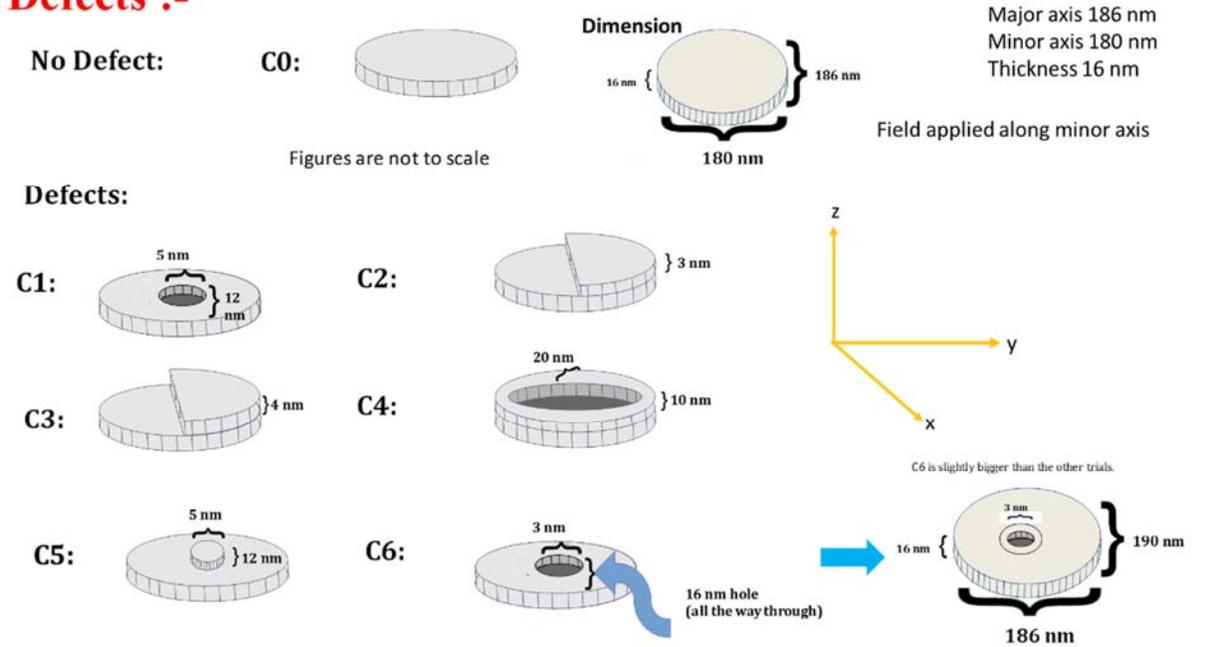

Fig. 2: Classification of defects in an elliptical cobalt nanomagnet of major axis dimension 186 nm, minor axis dimension 180 nm and thickness 16 nm: C0 (defect-free), C1 (hole in the center with diameter 5 nm and depth 12 nm), C2 (thickness variation where one half of the nanomagnet is 3 nm thicker, C3 (thickness variation where one half is 4 nm thicker), C4 (the periphery forms an annulus of width 20 nm and height 10 nm, C5 (thickness variation where a central circular region of diameter 5 nm is 12 nm thicker), and C6 (a through hole of 3 nm diameter at the nanomagnet's center; this nanomagnet dimension is slightly larger with major axis of 190 nm and minor axis of 186 nm). This figure is not drawn to scale.

## II. Excitation of spin waves in the defect-free and defective nanomagnets

Spin waves are generated in a magnet whenever its magnetization is perturbed by an external agent. They can be excited in a nanomagnet in a variety of ways. One common approach is to apply a bias magnetic field in the plane of the nanomagnet and then induce precession of the magnetization around this field with an ultrashort laser pulse. This is easily achieved in a time-resolved magneto-optical Kerr effect (TR-MOKE) and ferromagnetic resonance set-up. The precession spawns confined spin waves in the nanomagnet. In order to study them in the presence of defects, we simulate the following scenario: We consider cobalt nanomagnets in the form of elliptical disks whose major axis dimension is 186 nm, minor axis is 180 nm and thickness is 16 nm. A bias magnetic field is applied along the minor axis. Then an out-of-plane magnetic field pulse of amplitude 30 Oe, rise time 10 ps, and duration 100 ps is applied perpendicular to the nanomagnet's plane to simulate the effect of the laser pulse. This out-of-plane field sets the precession of the magnetization about the bias field in motion.

We track the time evolution of the nanomagnet's magnetization by using the micromagnetic simulator MuMax3[23] which allows us to determine the out-of-plane micromagnetic component $M_z(x,y,z,t)$ at every coordinate point within the nanomagnet at every instant of time. The nanomagnet is discretized into cells of dimension $2 \times 2 \times 2$ nm$^3$. The cell size in all directions is kept well below the exchange length of cobalt to consider both dipolar and exchange interactions in the magnetization dynamics of nanoscale magnets as well as to accurately reproduce the shapes of the nanomagnets under study. The time step used is 1 ps. The magnetic parameters used for the simulation are: saturation magnetization $M_s$= 1100 emu/cm$^3$, gyromagnetic ratio $\gamma$ = 17.6 MHz/Oe and exchange stiffness constant $A_{ex}$= 3.0 × 10$^{-6}$ erg/cm. These parameters correspond to cobalt nanomagnets. We spatially average $M_z(x,y,z,t)$ over space to find the out-of-plane magnetization component $\bar{M}_z(t)$ as a function of time.

We start the simulation by preparing the magnetic ground state upon applying the bias magnetic field ($H$) along the minor axis of the elliptical nanomagnet at time $t = 0$. The initial magnetization is assumed to have been directed along the major axis which is the easy axis. We wait until the micromagnetic distributions reach steady state and the spatially averaged magnetization points in the direction of the applied bias field $H$ along the minor axis. Next, we apply the out-of-plane magnetic field pulse and study the time evolution of the out-of-plane magnetization $\bar{M}_z(t)$ (associated with precession of the magnetization around the bias magnetic field) for 4 ns (4000 time steps).

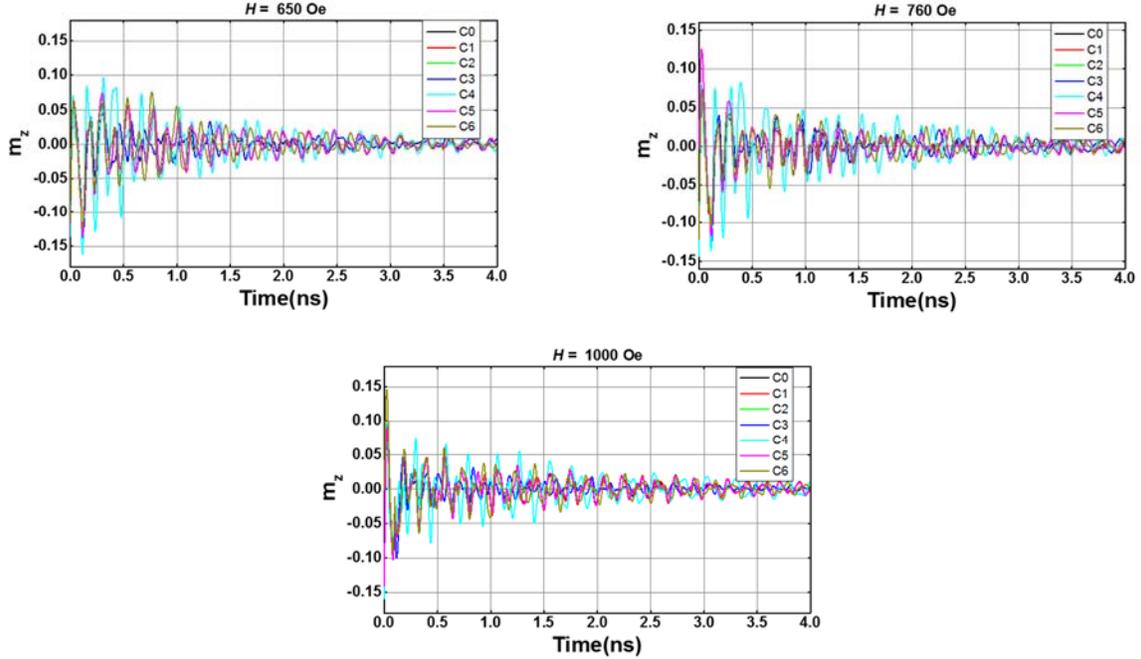

Fig. 3: The spatially averaged out-of-plane magnetization component as a function of time for the seven different (defect-free and defective) cobalt nanomagnets. The plots are for three different bias fields of strength 650 Oe, 760 Oe and 1000 Oe.

Fig. 3 shows $\bar{M}_z(t)$ versus $t$ for the seven different (defect-free and defective) nanomagnets at three different bias magnetic fields of strengths, $H$ = 650 Oe, 760 Oe and 1000 Oe. We perform a fast Fourier transform (FFT) of each of these "oscillations" to extract the dominant frequencies (frequency peaks) in the oscillation. These are the frequencies of the resonant spin wave modes in the nanomagnet. The frequency resolution in the generated FFT depends upon the total simulation time. Since the simulation time is 4 ns, the frequency resolution is 0.25 GHz.

Fig. 4 plots the Fourier spectra for the seven nanomagnets (nanomagnets with seven different types of defects illustrated in Fig. 2) at three different bias fields. The peaks in these spectra correspond to the frequencies of the resonant spin wave modes in the seven nanomagnets.

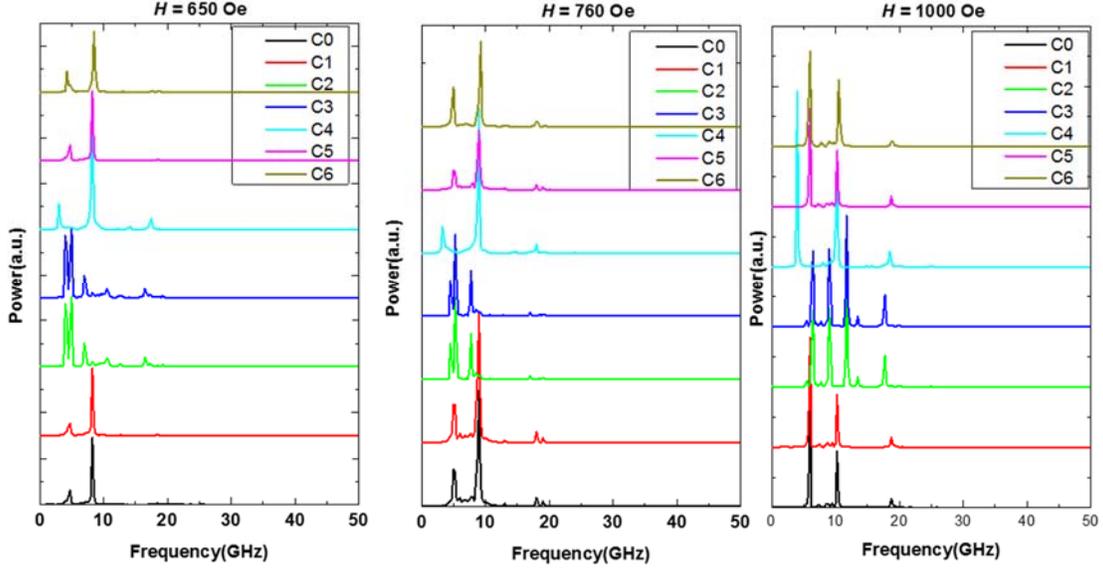

Fig. 4: Fast Fourier transforms of the $\bar{M}_z(t)$ versus $t$ oscillations for the seven different nanomagnets at three different bias field strengths ($H$). These are the frequencies of the dominant spin wave modes in the nanomagnets at the corresponding bias field.

There are three interesting features to note in Fig. 4. First, the spectral peaks, which are the frequencies of the resonant spin wave modes, are different in the seven different nanomagnets at the same bias magnetic field. This shows that *defects affect the frequencies of the resonant spin wave modes* and that has important implications for spin wave microwave generators. Second, the bias field dependence of the resonant mode frequencies (peaks in the spectrum) are sensitive to defects. Thus, the "tunability" of the oscillation frequency of microwave oscillators with a magnetic field is affected by the presence of defects. Third, defects spawn some new resonant modes that are absent in the defect-free nanomagnet. Conversely, defects can also quench resonant modes that are present in the defect-free nanomagnets.

### III.     Spin wave modes: power and phase profiles

In order to calculate the power and phase profiles of the spin wave modes, we proceed as follows: We fix the z-coordinate at a particular value $(z = z_m)$ and perform a discrete Fourier transform (FFT) of $M(x, y, z, t)|_{z=z_m}$ (obtained from MuMax3 simulation) with respect to time to yield $\tilde{M}^{z_m}(f, x, y) = FFT\left[ M^{z_m}(t, x, y, z)|_{z=z_m} \right]$. The fixed value of the z-coordinate $(z_m)$ is chosen to be at the surface of the nanomagnet. Our in-house software *Dotmag* then plots the $(x, y)$ spatial distribution of the power and phase of the spin waves at chosen frequencies $f_n$ on the surface of the nanomagnet $(z = z_m)$ according to the following relations[24,25].

$$P^{z_m, f_n}(x, y) = 20 \log_{10} \tilde{M}^{z_m}(f_n, x, y);$$

$$\phi^{z_m, f_n}(x, y) = \tan^{-1}\left(\frac{\text{Im}(\tilde{M}^{z_m}(f_n, x, y))}{\text{Re}(\tilde{M}^{z_m}(f_n, x, y))}\right), \quad (1)$$

where $f_n$ is the frequency of a resonant mode. The power is expressed in dB and the minimum value of $\tilde{M}^{z_m}(f_n, x, y)$ is normalized to unity.

## IV.     Results and Discussion

Figures 5-11 show the power and phase profiles of the resonant spin wave modes in the seven different nanomagnets. We observe four different types of modes: center mode (where power is localized at the center of the nanomagnet), edge mode (where power is localized at the edges), quantized mode (where power is localized at discrete locations) and mixed quantized mode. For some mixed quantized mode, it is difficult to define an exact quantization number owing to the blurred and equivocal nature of the profile.

In a continuous thin film, there are three types of magnetostatic spin-waves. They are categorized based upon the relative orientation of the magnetization (***M***) and the wave vector (***k***) of the spin wave.

I) <u>Damon-Eshbach (DE) mode</u> – In this mode magnetization and the wave vector both lie in the film plane and form an angle $\phi \approx 90°$. This is also known as magnetostatic surface wave (MSSW) mode[26].

II) <u>Backward Volume (BV) mode</u> - In this mode, the magnetization and the wave vector both are collinear and lie in the film plane[27].

III) <u>Forward Volume (FV) mode</u> - In the so-called magnetostatic forward volume mode (MSFVM) geometry, the magnetization is perpendicular to the film[28].

In our system of nanomagnets, modes that are similar to the first two types of modes can exist since we apply the bias magnetic field in the plane of the sample. Nothing like the third type of mode can exist. Since the modes in the nanomagnets are not propagating modes, they are not exactly classifiable as DE or BV modes. The generated spin waves get reflected from the boundaries of the nanomagnets and form standing spin-wave modes similar to resonant cavity modes. Hence, we call them resonant modes. Because of their confined nature, instead of assigning wave vector to the modes, we count the number of nodal planes and assign a mode quantization number to the observed modes. The quantization numbers are defined according to whether the quantization axis is along the magnetic field direction (*n*, in BV geometry) or perpendicular to the field direction (*m'*, in DE geometry). In some cases, the modes are quantized along the azimuthal axis. We name those as azimuthal modes with a corresponding quantization number (*m*).

From Figures 5-11, we see that certain types of defects (C1 and C5) are relatively innocuous and affect the resonant spin wave modes slightly. They do not spawn new modes or quench existing ones. The changes they introduce in the power and phase profiles are also moderate. These types of defects are very tolerable.

Defect C6 is similar to C1, but unlike C1, this is a through-hole which makes it more invasive (the size is also slightly larger). The through-hole spawns a new mode at 650 Oe bias magnetic field and quenches an existing mode at 760 Oe field. It does not alter the power profiles significantly, but affects the phase profiles much more. These types of defects are moderately tolerable, except in applications that require phase sensitivity.

Defects C2 and C3 are associated with thickness variation across one-half of the nanomagnet's surface. These types of defect are found to be extremely invasive and spawn new resonant modes at all magnetic fields. They also alter the power and phase profiles of the resonant modes quite significantly. Thickness variation across a significant fraction of a nanomagnet's surface (an extended defect) is therefore more serious than having localized defects such as a "hole" (C1, C6) or a "hillock" (C5). These types of defects are found to be the most harmful among the ones studied.

Defect C4 is important since it is commonplace in nanomagnets fabricated by electron-beam evaporation of a ferromagnetic metal into a lithographically delineated window in an e-beam resist. Curiously, it is not as invasive as C2 and C3. Like C6, it changes the power profiles slightly, but affects the phase profiles much more. The frequency of the edge mode decrease significantly. It also quenches a quantized mode that appears in the defect-free nanomagnet at the intermediate field of 760 Oe. This type of defect is, again, moderately tolerable, except in applications that hinge on phase sensitivity.

Based on these observations, it appears that maintaining thickness uniformity across a significant fraction of the nanomagnet's surface would be critical in applications that require reproducibility of resonant spin wave power and phase profiles. Expectedly, extended defects have a more serious effect on the resonant spin wave power and phase profiles than localized defects.

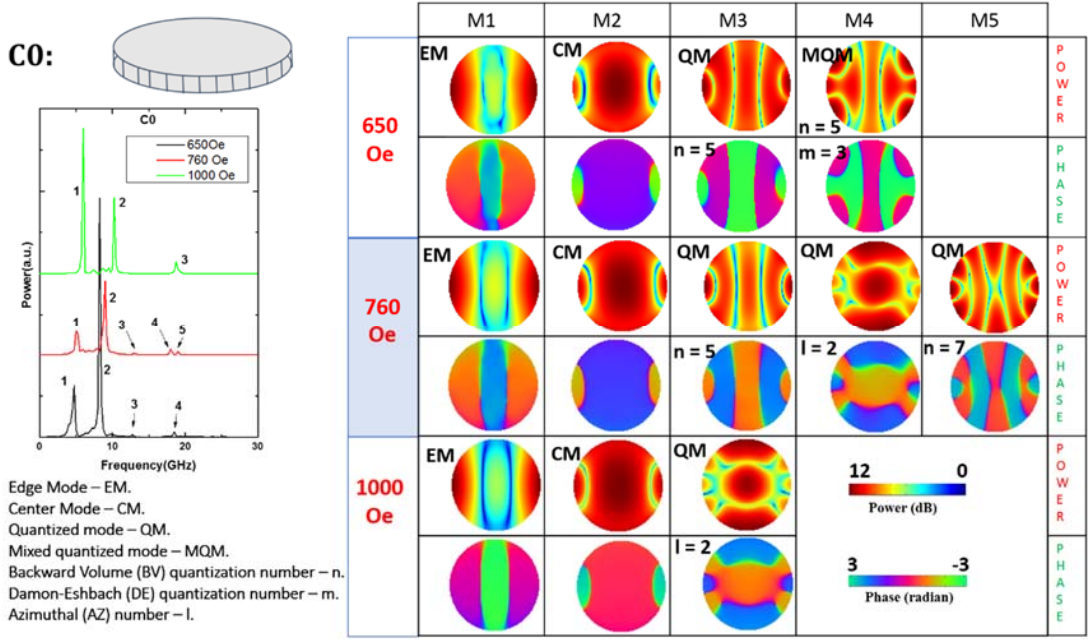

Fig. 5: Power (top row) and phase (bottom row) profiles of the resonant spin wave modes on the surface of the defect-free nanomagnet (C0) at three different bias magnetic fields. The left panel shows the Fourier transform spectra $\tilde{M}^{z_m}(f,x,y)$ at the three different fields in the defect-free nanomagnet and the right panel shows the power and phase distributions in space (profiles). The resonant modes are the ones that correspond to the peak frequencies in the Fourier transform spectra shown in the left panel and are labeled M1, M2, …These modes are either center mode, edge mode, quantized mode or mixed quantized mode. The phase varies from $-\pi$ ($\approx -3$) to $\pi$ ($\approx 3$) and the power varies from 0 to 12 dB.

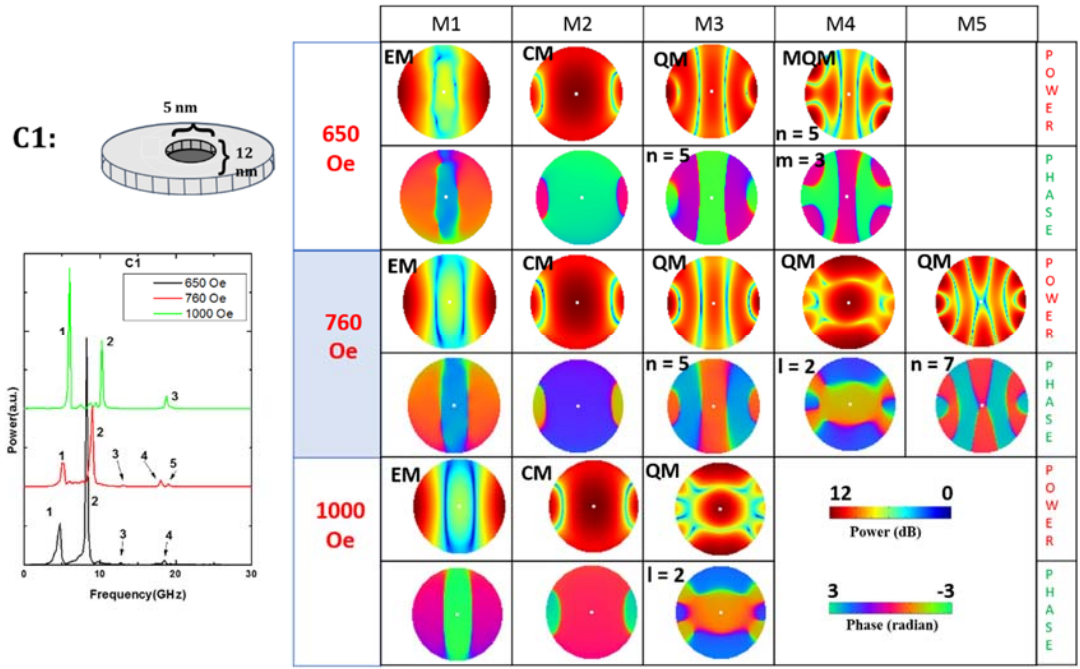

Fig. 6: Power (top row) and phase (bottom row) profiles of the resonant spin wave modes on the surface of the nanomagnet with defect of type C1 at three different bias magnetic fields. No new mode is spawned and none is quenched by the defect.

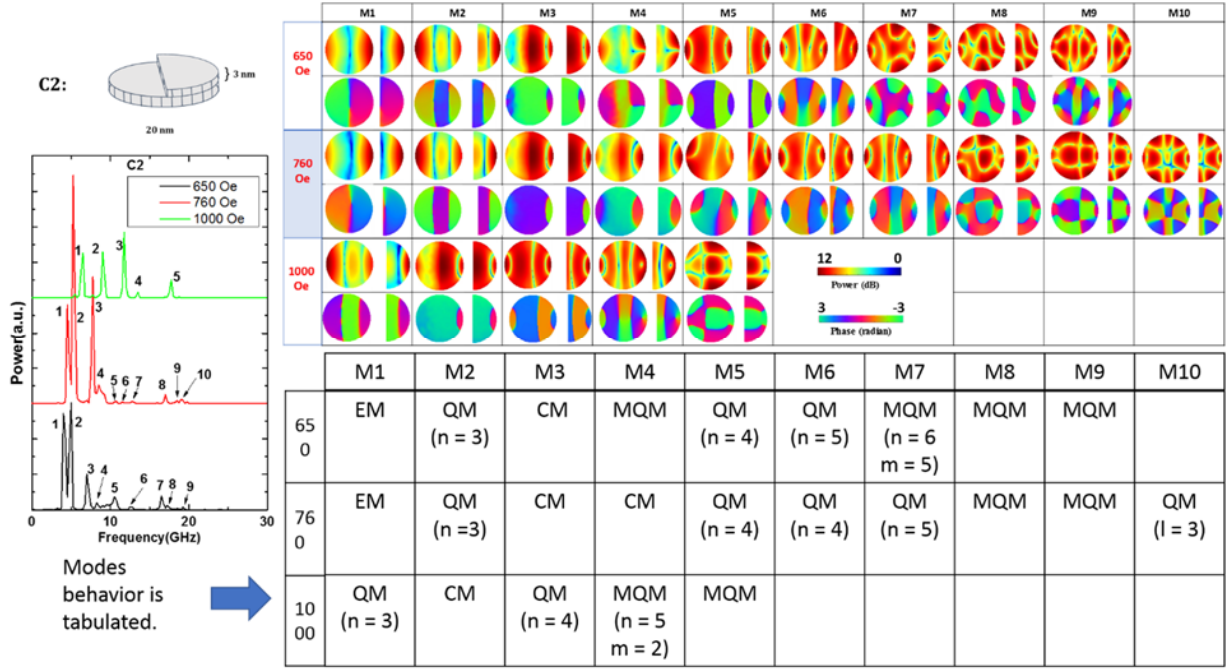

Fig. 7: Power and phase profiles of the resonant spin wave modes on the surface of the nanomagnet with defect C2 at three different bias fields. The top row shows the power profile and the bottom row the phase profile at any given bias field strength. The left panel shows the Fourier transform spectra $\tilde{M}^{z_m}(f,x,y)$ at three different bias magnetic fields. There are a large number of resonant modes in this case (magnet thickness varies), many of which are absent in the defect-free nanomagnet. Clearly, this type of defect spawns new modes.

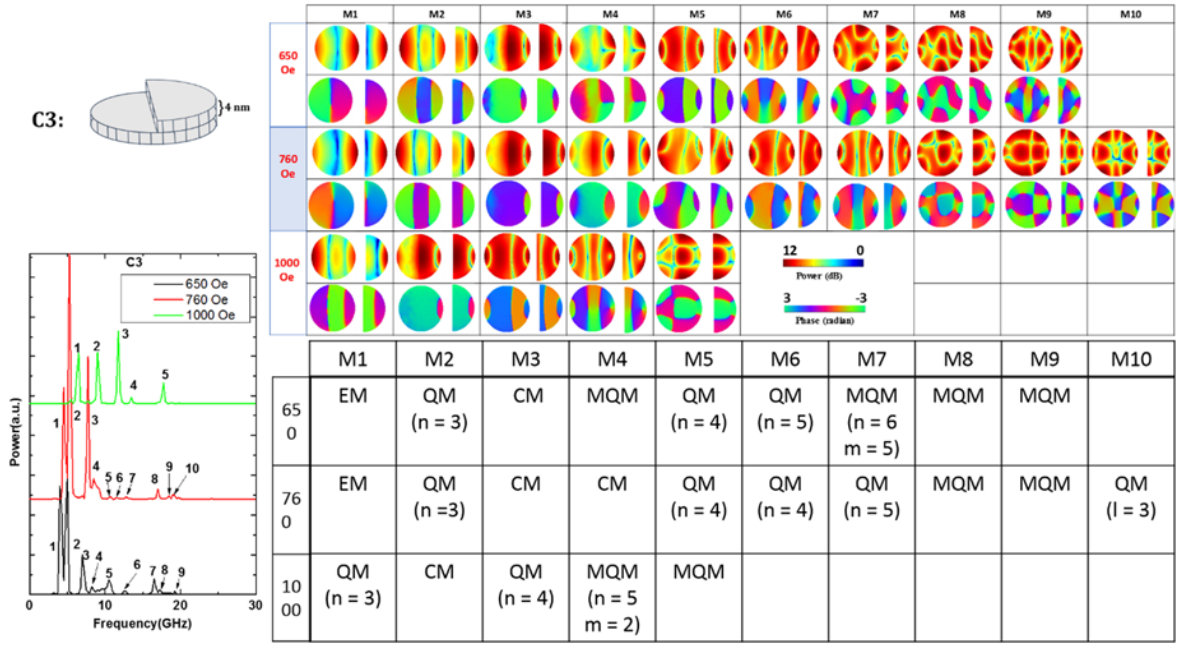

Fig. 8: Power (top row) and phase (bottom row) profiles of the resonant spin wave modes on the surface of the nanomagnet with defect C3 at three different bias fields. The left panel shows the Fourier transform spectra $\tilde{M}^{z_m}(f,x,y)$ at three different bias magnetic fields. New resonant modes are spawned by this defect.

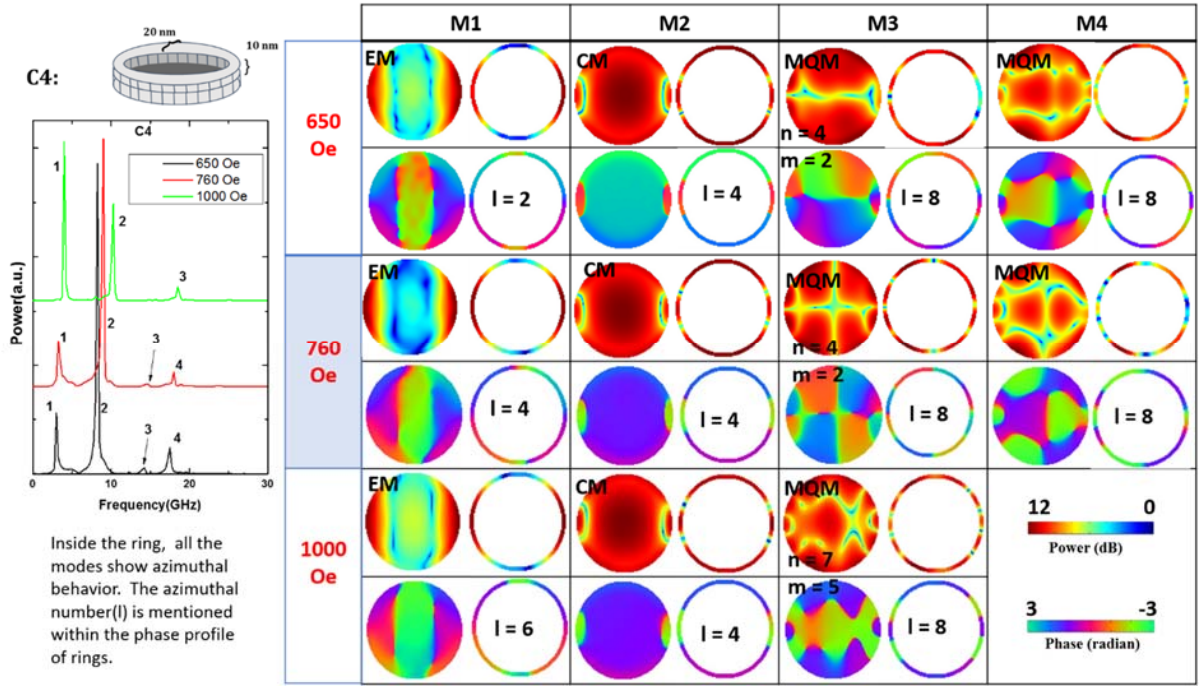

Fig. 9: Power (top row) and phase (bottom row) profiles of the resonant spin wave modes on the surface of the nanomagnet with defect C4 at three different bias magnetic fields. This is the case that most closely approximates the structure in Fig. 1. Here, we show, separately, the power and phase profiles in the surface of the nanomagnet (left) and in the surface of the 20 nm thick annulus (right). No new mode is spawned, but one is quenched.

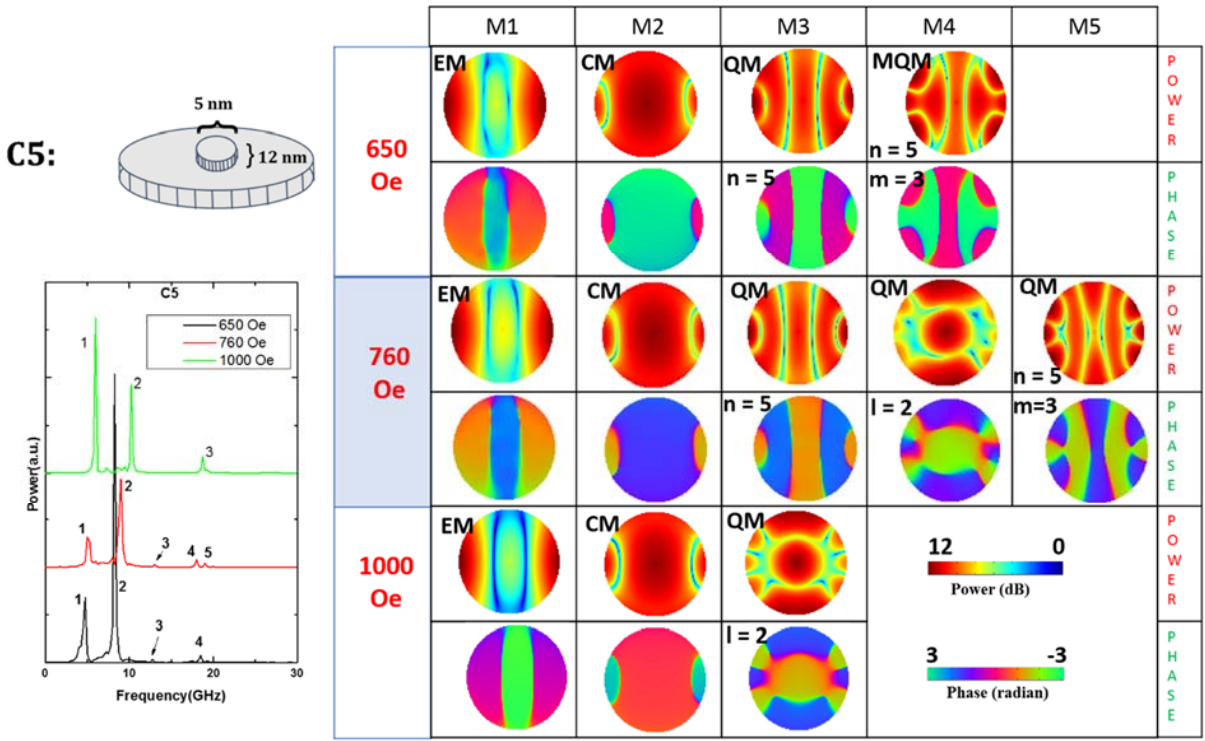

Fig. 10: Power (top row) and phase (bottom row) profiles of the resonant spin wave modes on the surface of the nanomagnet C5 at three different magnetic fields. No new mode is spawned and none quenched.

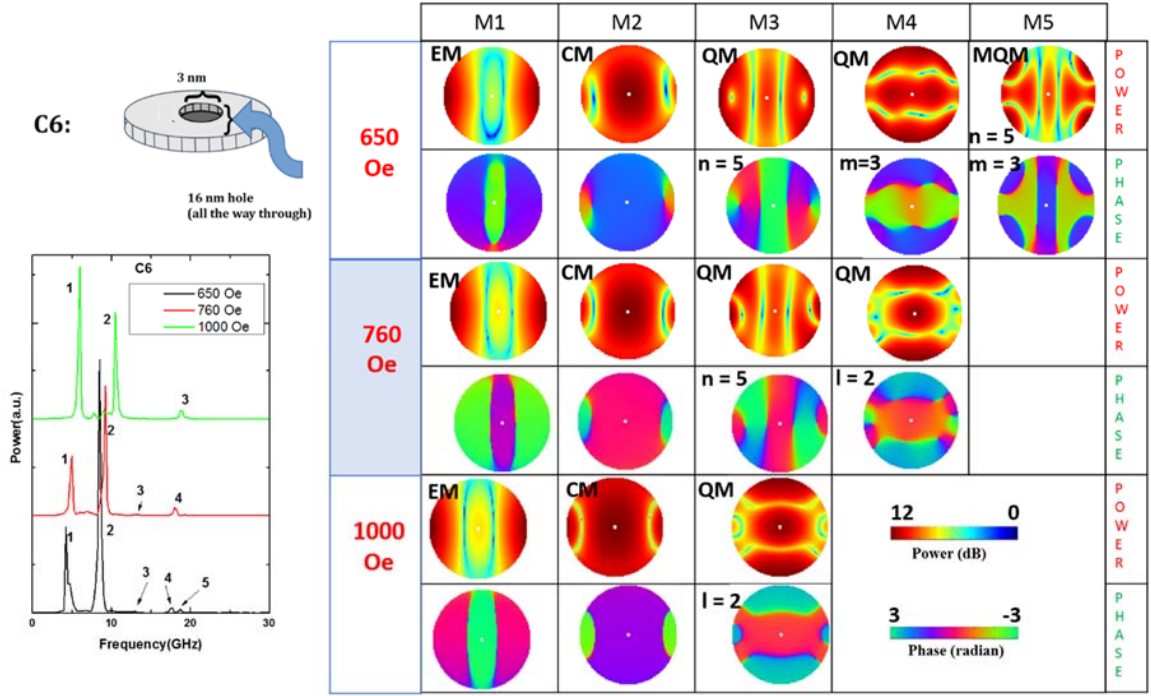

Fig. 11: Power (top row) and phase (bottom row) profiles of the resonant spin wave modes on the surface of the nanomagnet C6 at three different magnetic fields. A new mode is spawned at the low field and one is quenched at the intermediate field.

## V. CONCLUSION

Our study has shown that several features of resonant spin wave modes in a nanomagnet (frequencies, magnetic field dependence of the frequencies, number and nature of resonant modes, power and phase profiles) are affected by the presence of defects associated with localized or extended thickness variations in the plane of the nanomagnets. This has serious consequences for many applications that rely on spin wave modes. In the past, it was found that in magnetostrictive nanomagnets, strain-induced magnetization reversal (switching) probability is dramatically affected by the presence of defects[29, 30]. Defects are also known to have a serious deleterious effect on the stochastic behavior of low energy barrier nanomagnets that have been proposed for use in stochastic computing[31]. Here, we have found that defects have a dramatic effect on spin wave modes as well. For example, C1 and C6 are slightly different defects and yet the spin wave modes are vastly different in them. This indicates that spin waves are very sensitive to defects and that will cause significant device-to-device variability since the defect morphology will be different in different devices. In single (or few) device applications like magnonic holography[32] or a magnonic gate[33] or spin wave interferometer[34] or modulator[35], this will not matter much since the number of devices involved is one or few, but in large-scale spin wave "circuits" where numerous devices have to behave in nominally identical manner for overall circuit functionality, the device-to-device variability caused by defects could be debilitating. Spin wave circuits that have little tolerance for variations of spin wave frequencies, or their power distributions, or their phase profiles – e.g. phase locked nano-oscillators for neuromorphic computing[5,22] – are especially vulnerable. Designing these systems for targeted applications in the presence of random defects will be extremely challenging.

**Acknowledgements:**

The work at Virginia Commonwealth University was supported by the US National Science Foundation under grants ECCS-1609303 and CCF-1815033. The work at S. N. Bose National Center for Basic Sciences was supported by the Center (grant No. SNB/AB/18-19/211). S. S. acknowledges S. N. Bose National Center for Basic Sciences for Senior Research Fellowship.

**Author Contributions**

M. A. A. and S. S. did all the simulations, M. A. A. fabricated the nanomagnets in Fig. 1 and characterized them with atomic force microscopy, A. B. and S. B. designed and oversaw the research. All authors contributed to writing the paper.

The authors declare no competing interest.

All data in the manuscript are available from the authors upon written request and subject to restrictions imposed by the authors' institutions and funding agencies.

# Supplementary Material: The effect of material defects on resonant spin wave modes in a nanomagnet

Md Ahsanul Abeed[1], Sourav Sahoo[2], David Winters[1], Anjan Barman[2] and Supriyo Bandyopadhyay[1]

[1]Dept. of Electrical and Computer Engineering, Virginia Commonwealth University, Richmond, VA 23284, USA

[2]Dept. of Condensed Matter Physics and Materials Science, S. N. Bose National Center for Basic Sciences, Kolkata 700 116, INDIA

In this supplementary material, we show the steady-state micromagnetic distributions within the six defective nanomagnets and the defect-free nanomagnet at the bias magnetic field of 650 Oe. The distributions at higher magnetic fields are not qualitatively different and hence not shown.

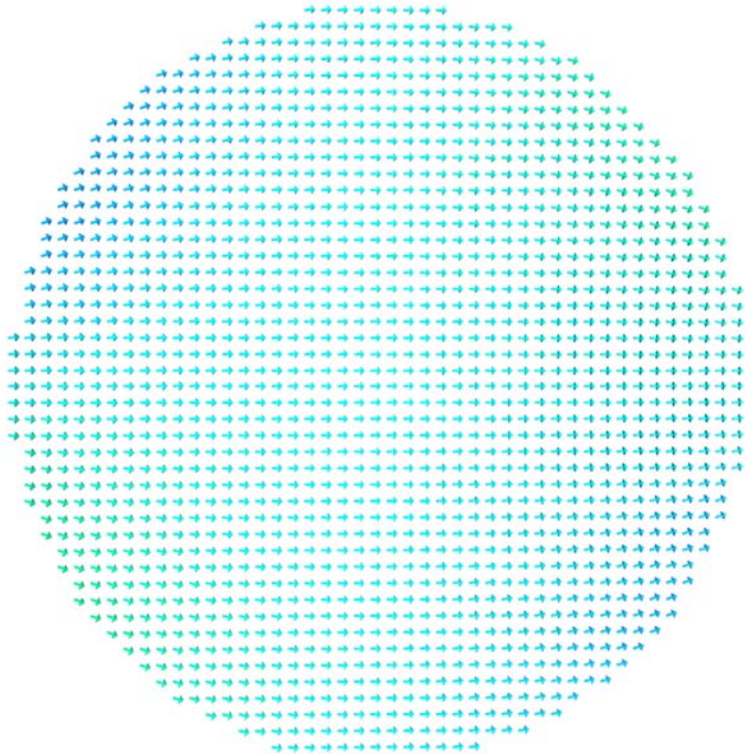

Fig. S1: Steady state micromagnetic distributions in the top layer of the defect-free nanomagnet (C0). The magnetic field is 650 Oe and is directed to right along the horizontal axis.

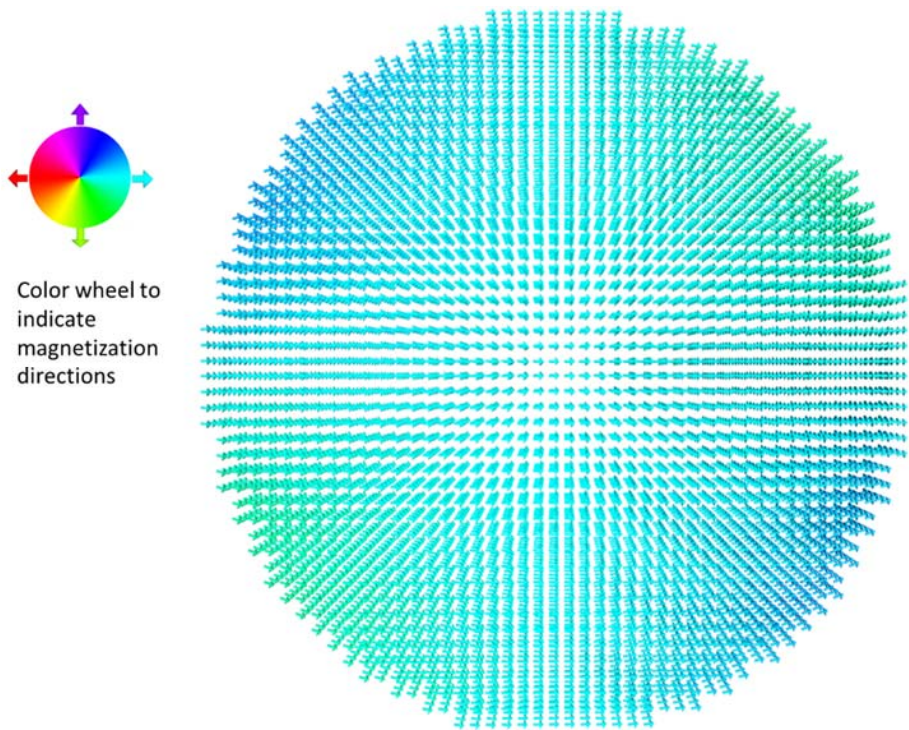

Fig. S2: Steady state micromagnetic distributions in all layers of the defect-free nanomagnet (C0). The magnetic field is 650 Oe and is directed to right along the horizontal axis.

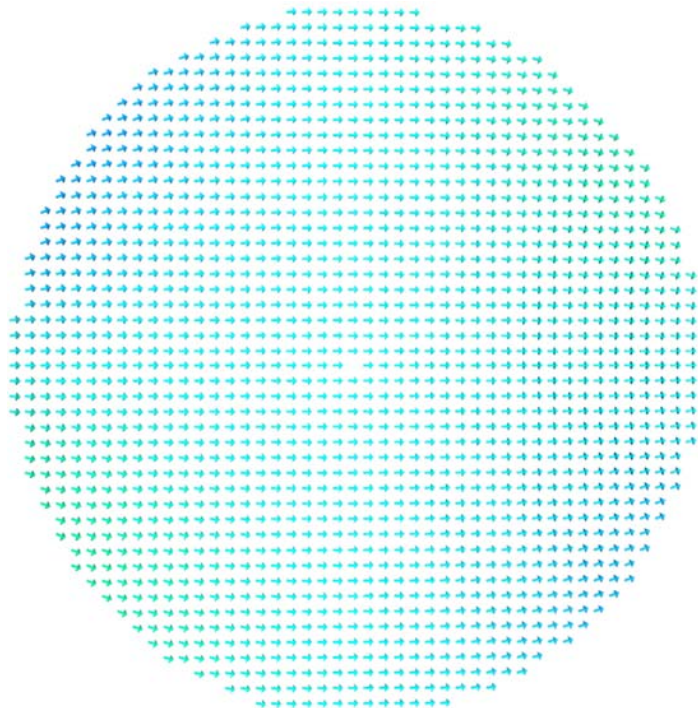

Fig. S3: Steady state micromagnetic distributions in the top layer of the defective nanomagnet C1. The magnetic field is 650 Oe and is directed to right along the horizontal axis.

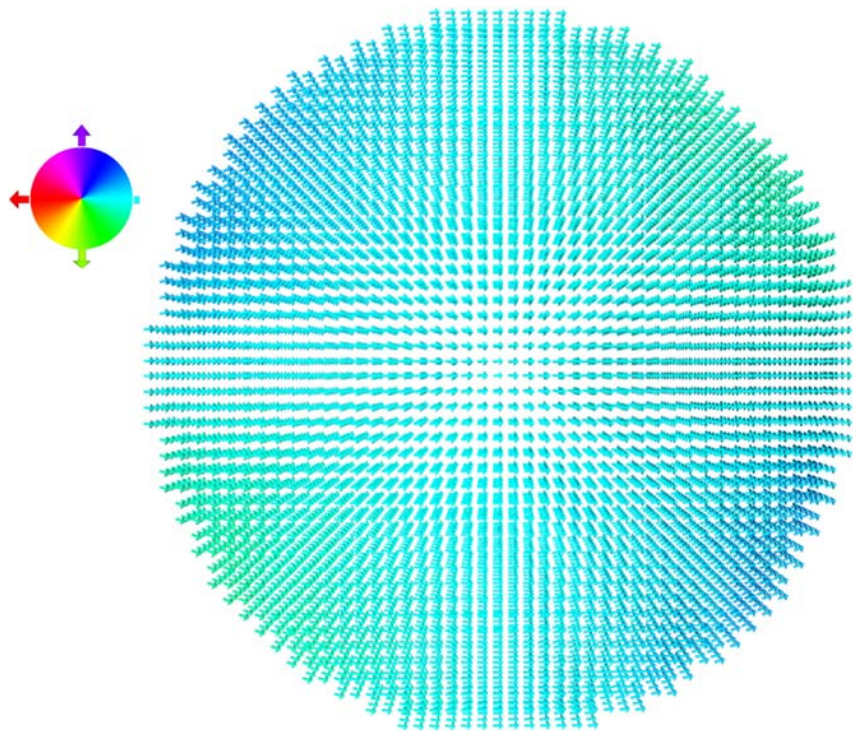

Fig. S4: Steady state micromagnetic distributions in all layers of the defective nanomagnet C1. The magnetic field is 650 Oe and is directed to right along the horizontal axis.

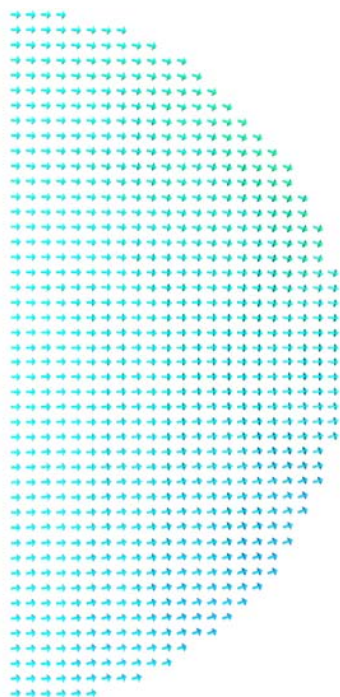

Fig. S5: Steady state micromagnetic distributions in the top layer of the defective nanomagnet C2. The magnetic field is 650 Oe and is directed to right along the horizontal axis.

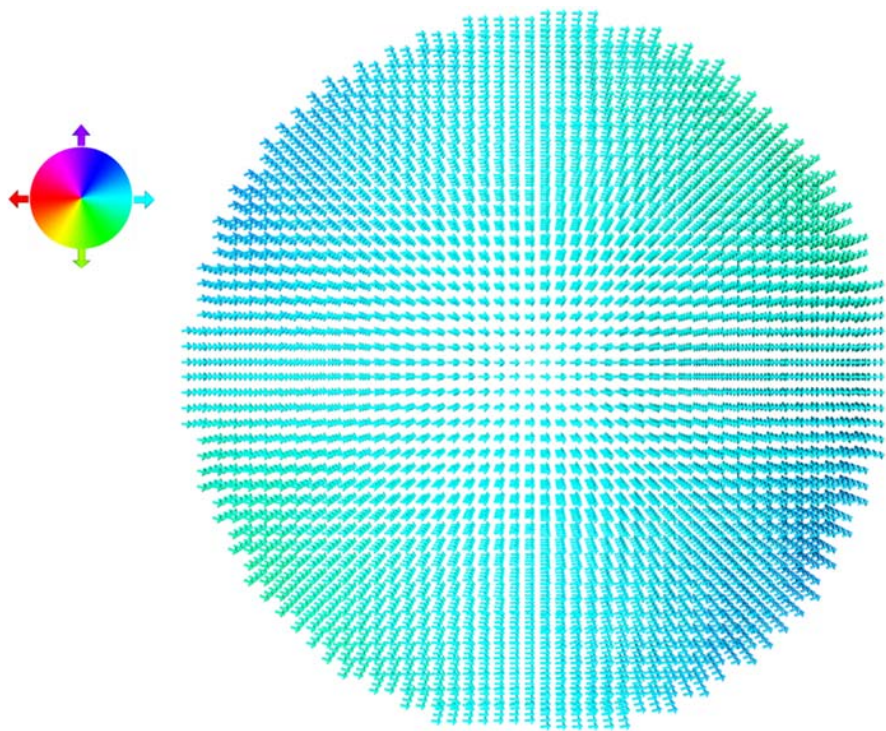

Fig. S6: Steady state micromagnetic distributions in all layers of the defective nanomagnet C2. The magnetic field is 650 Oe and is directed to right along the horizontal axis.

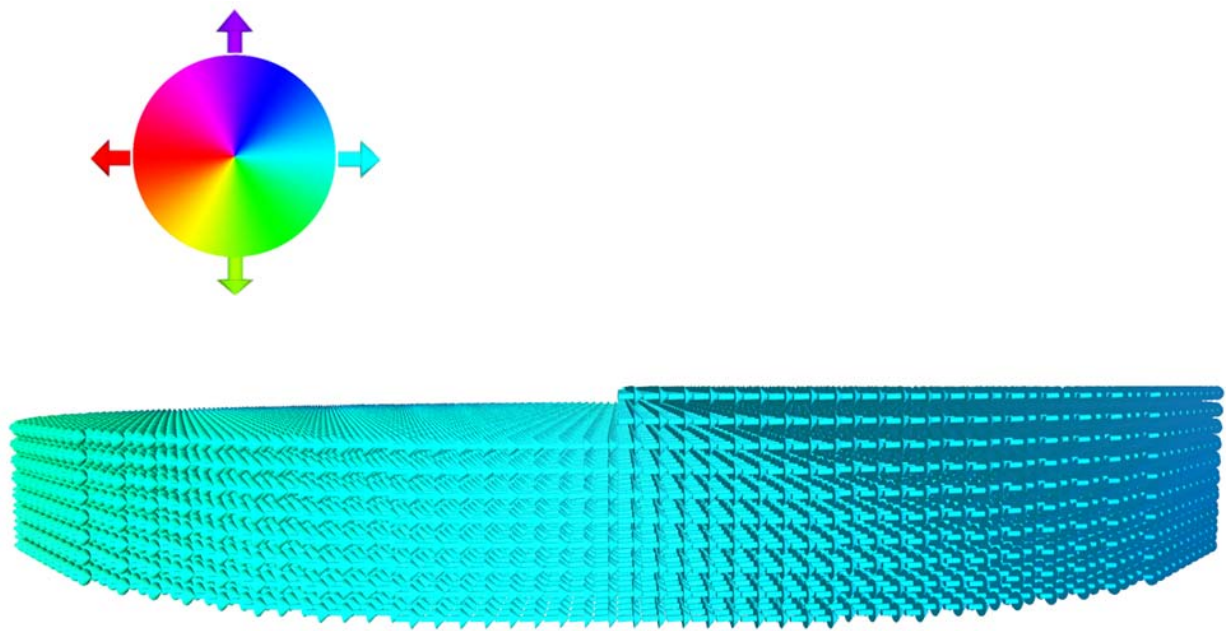

Fig. S7: Side view of the steady state micromagnetic distributions in all layers of the defective nanomagnet C2. The magnetic field is 650 Oe and is directed to right along the horizontal axis.

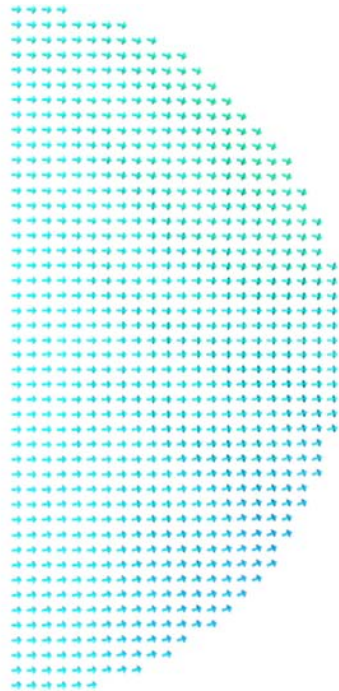

Fig. S8: Steady state micromagnetic distributions in the top layer of the defective nanomagnet C3. The magnetic field is 650 Oe and is directed to right along the horizontal axis.

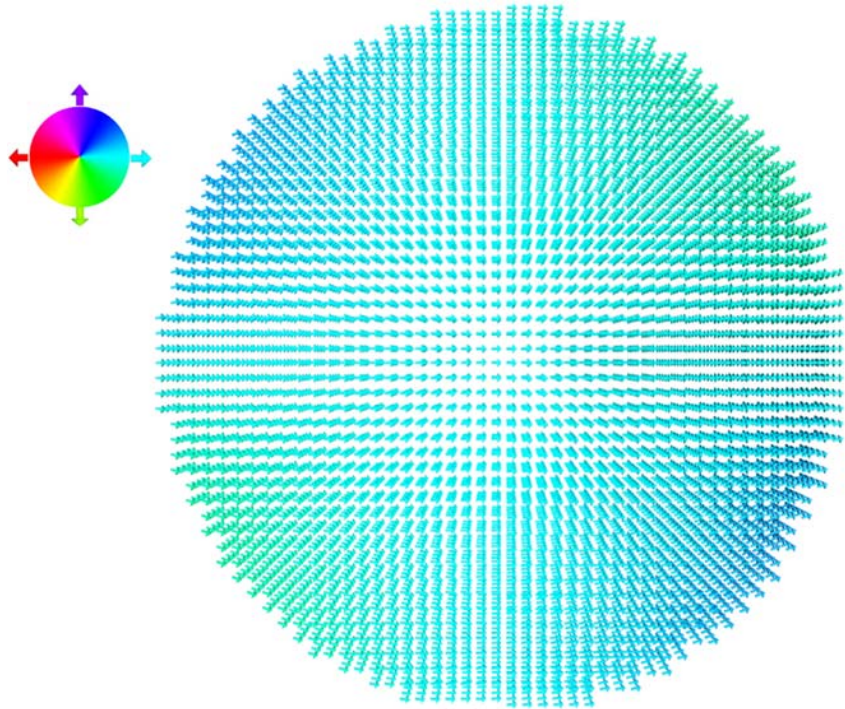

Fig. S9: Steady state micromagnetic distributions in all layers of the defective nanomagnet C3. The magnetic field is 650 Oe and is directed to right along the horizontal axis.

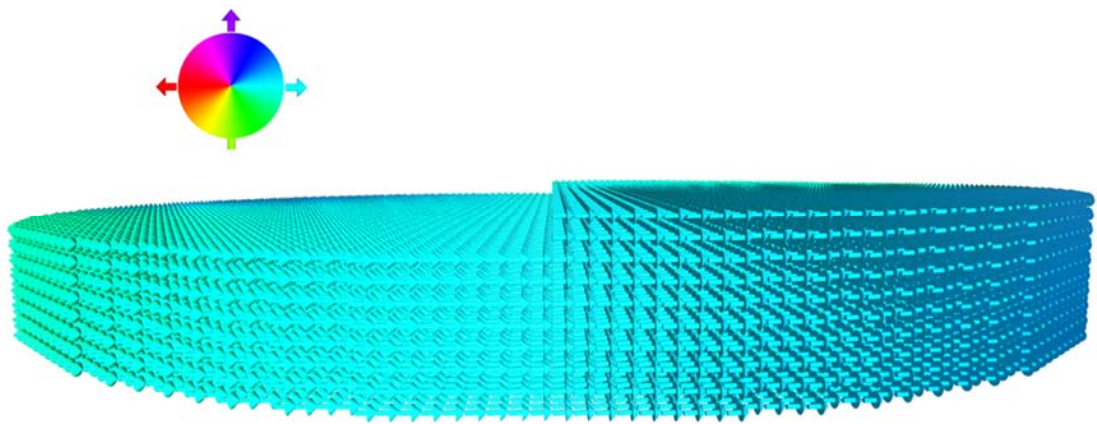

Fig. S10: Side view of the steady state micromagnetic distributions in all layers of the defective nanomagnet C3. The magnetic field is 650 Oe and is directed to right along the horizontal axis.

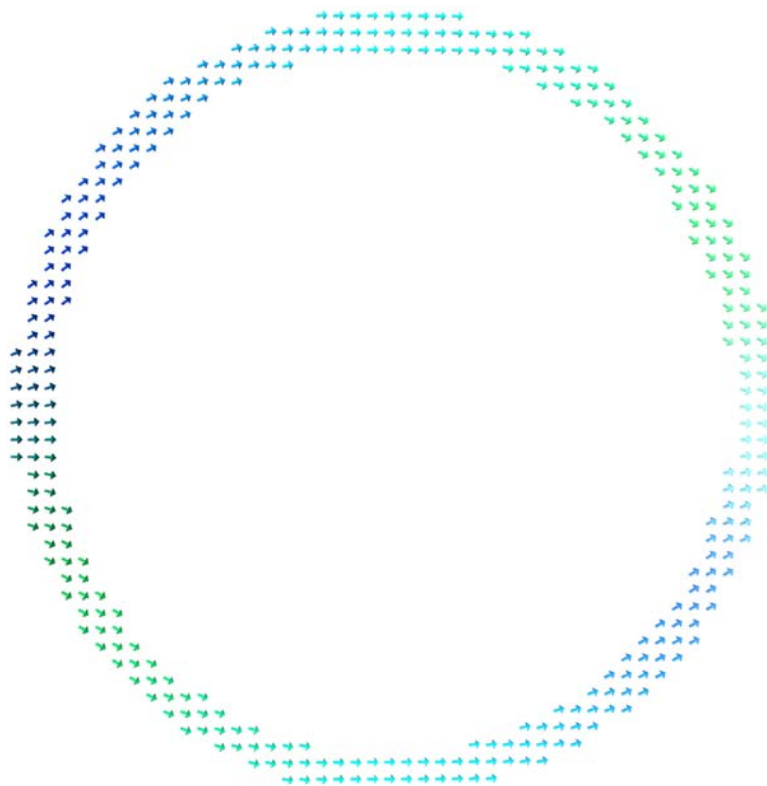

Fig. S11: Steady state micromagnetic distributions in the top layer of the defective nanomagnet C4. The magnetic field is 650 Oe and is directed to right along the horizontal axis.

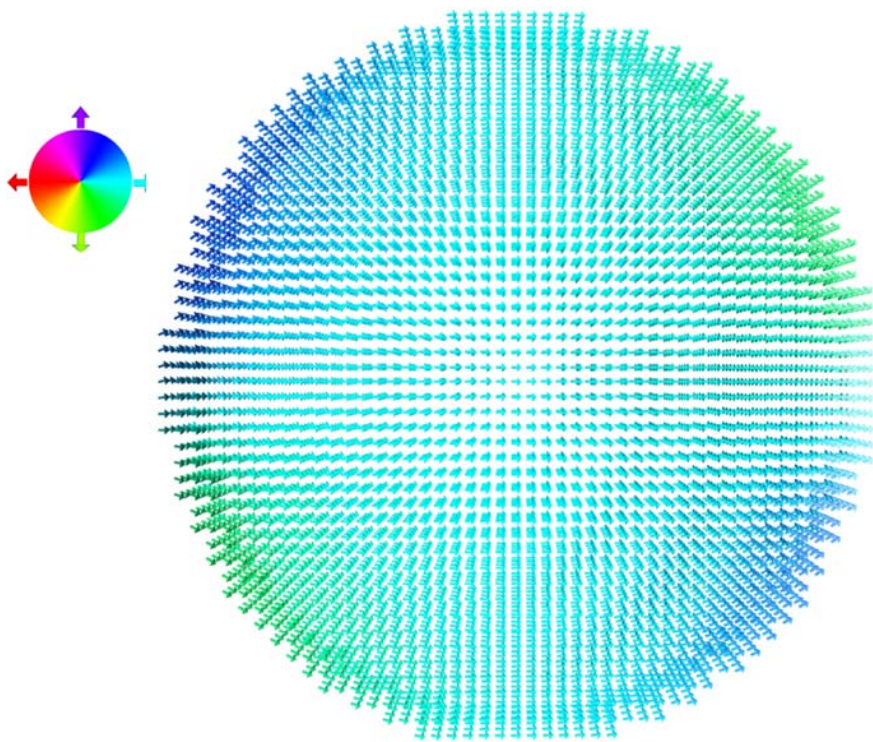

Fig. S12: Steady state micromagnetic distributions in all layers of the defective nanomagnet C4. The magnetic field is 650 Oe and is directed to right along the horizontal axis.

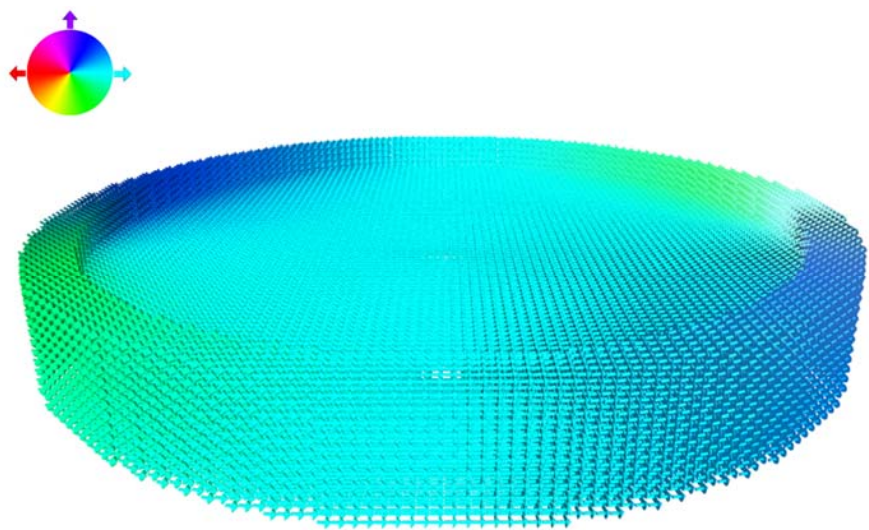

Fig. S13: Side view of the steady state micromagnetic distributions in all layers of the defective nanomagnet C4. The magnetic field is 650 Oe and is directed to right along the horizontal axis.

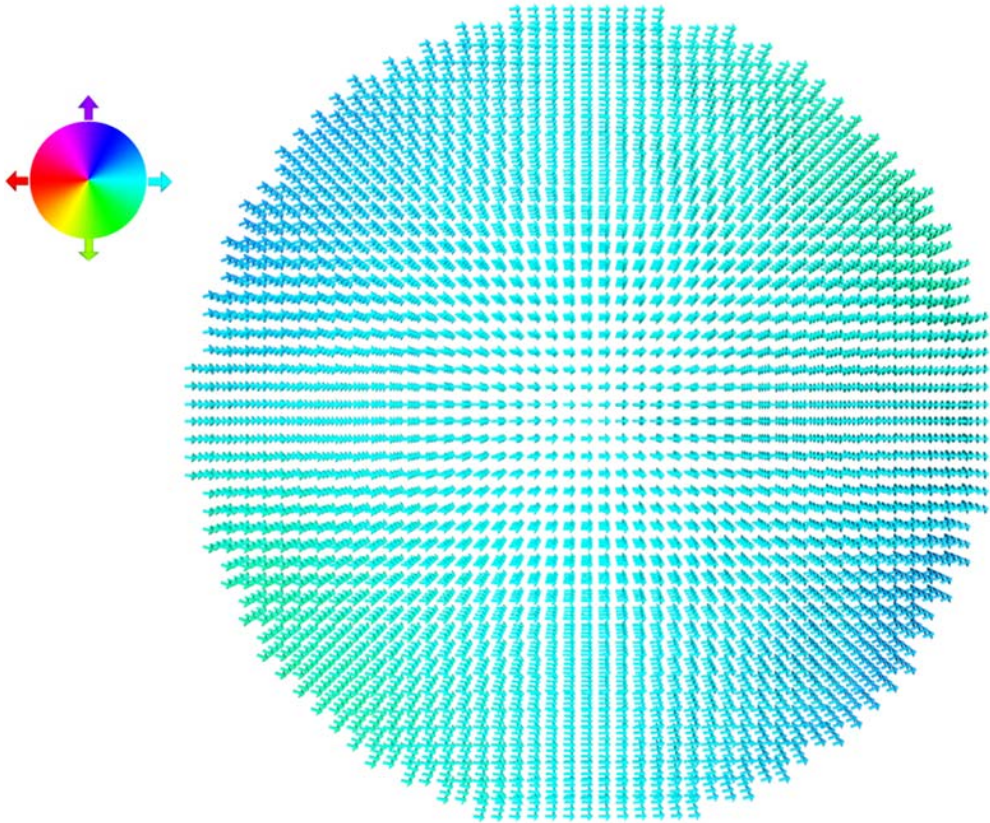

Fig. S14: Steady state micromagnetic distributions in all layers of the defective nanomagnet C5. The magnetic field is 650 Oe and is directed to right along the horizontal axis.

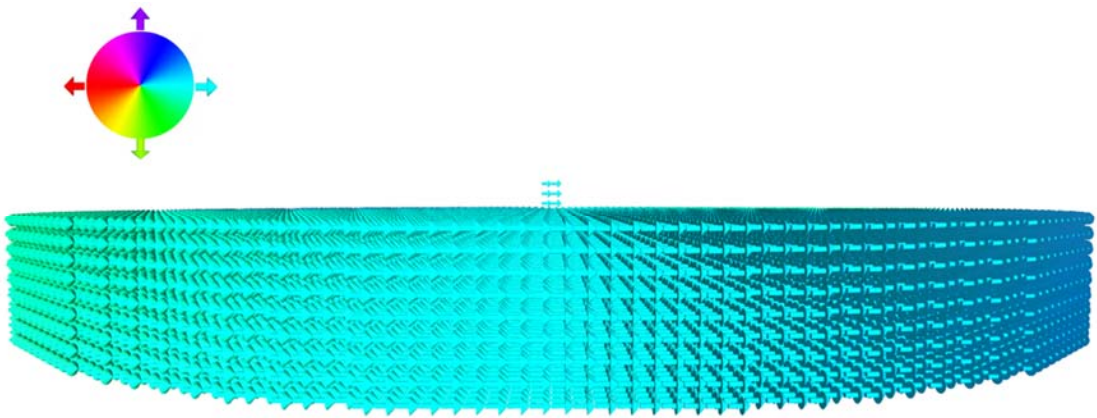

Fig. S15: Side view of the steady state micromagnetic distributions in all layers of the defective nanomagnet C5. The magnetic field is 650 Oe and is directed to right along the horizontal axis.

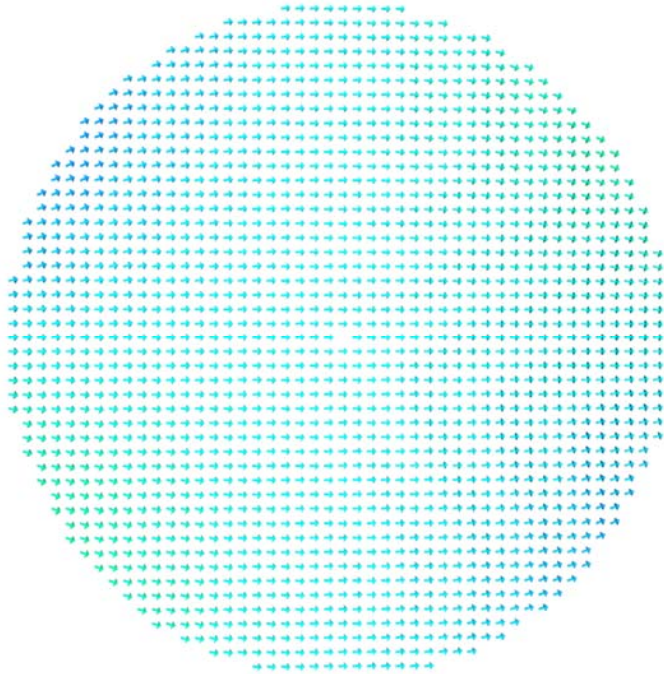

Fig. S16: Steady state micromagnetic distributions in the top layer of the defective nanomagnet C6. The magnetic field is 650 Oe and is directed to right along the horizontal axis.

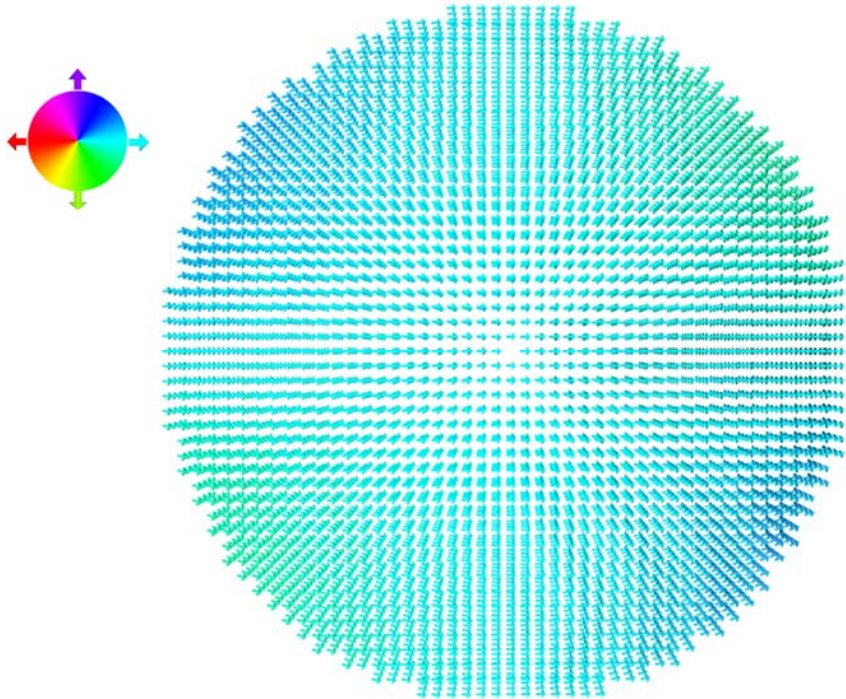

Fig. S17: Steady state micromagnetic distributions in all layers of the defective nanomagnet C6. The magnetic field is 650 Oe and is directed to right along the horizontal axis.